# An Integrated Deep Learning Framework Leveraging NASNet and Vision Transformer with MixProcessing for Accurate and Precise Diagnosis of Lung Diseases

Sajjad Saleem[1]
Department of Information and Technology, Washington University of Science and Technology, Alexandria, VA 22314, USA.
ssaleem.student@wust.edu

Muhammad Imran Sharif[2*]
Department of Computer Science, Kansas State University, Manhattan, KS 66506, USA.

imransharif@ksu.edu

Corresponding Author:
Muhammad Imran Sharif
imransharif@ksu.edu

## Abstract

The lungs are the essential organs of respiration, and this system is significant in the carbon dioxide and exchange between oxygen that occurs in human life. However, several lung diseases, which include pneumonia, tuberculosis, COVID-19, and lung cancer, are serious healthiness challenges and demand early and precise diagnostics. The methodological study has proposed a new deep learning framework called NASNet-ViT, which effectively incorporates the convolution capability of NASNet with the global attention mechanism capability of Vision Transformer ViT. The proposed model will classify the lung conditions into five classes: Lung cancer, COVID-19, pneumonia, TB, and normal. A sophisticated multi-faceted preprocessing strategy called MixProcessing has been used to improve diagnostic accuracy. This preprocessing combines wavelet transform, adaptive histogram equalization, and morphological filtering techniques. The NASNet-ViT model performs at state of the art, achieving an accuracy of 98.9%, sensitivity of 0.99, an F1-score of 0.989, and specificity of 0.987, outperforming other state of the art architectures such as MixNet-LD, D-ResNet, MobileNet, and ResNet50. The model's efficiency is further emphasized by its compact size, 25.6 MB, and a low computational time of 12.4 seconds, hence suitable for real-time, clinically constrained environments. These results reflect the high-quality capability of NASNet-ViT in extracting meaningful features and recognizing various types of lung diseases with very high accuracy. This work contributes to medical image analysis by providing a robust and scalable solution for diagnostics in lung diseases.

## 1. Introduction

Deep learning algorithms have revolutionized the face of medical image analysis and are very promising for detecting and classifying lung diseases. This introduction draws on key research articles that explain the applications of the use of DL models to classify and diagnose different lung conditions, including Lung cancer, COVID-19, pneumonia, TB, and normal. One of the major contributions in this area came from Wang et al. [1] when they made the ChestX-ray8 dataset available, which is crucial for developing various DL models needed for chest X-ray image analysis. The dataset provided a basis for the diagnosis of pneumonia and other lung anomalies, with standards supporting the weakly supervised annotation and localization of common thoracic diseases. In parallel, CNNs have shown impressive performance in diagnosing lung diseases. Multi-scale dense networks were introduced by Shen et al. [2], which leverage deep CNN architectures to enhance image classification. This has been consistently effective in the hierarchical feature extraction for accurately identifying lung diseases. COVID-19 underlined the need for rapid and efficient diagnostic tools. Li et al. [3] proposed a different AI system for COVID-19 identification from community-obtained pneumonia in chest CT images. By using DL methods, this method helps find COVID-19-positive cases and their discrimination, helping better manage and control the disease. TB is another major lung disease and has also been one of the major points of extensive study regarding DL methods. Lakhani and Sundaram [4] suggested a deep learning model for automatically classifying chest radiographs for pulmonary tuberculosis. The proposed network accurately detected the presence of TB-related abnormalities and, therefore, was useful for effective screening and diagnosis. Their network also demonstrated great potential in identifying and characterizing various pattern variations related to interstitial lung diseases for further appropriate identification and medication. The use of DL for COVID-19 detection has drawn a lot of attention lately. Apostolopoulos and Mpesiana [5] detected COVID-19 from X-ray pictures by utilizing CNNs and transfer learning. Their approach indicated that DL could be helpful in the early detection of patients who suffer from COVID-19 by supporting the medical diagnosis made by radiologists and healthcare professionals and helpful for healthcare responders during the pandemic. Anthimopoulos et al. [6] proposed a deep CNN for classifying lung patterns limited to interstitial lung disorders. It was a significant turning point in the categorization of lung diseases. Similarly, Jin et al. [7] designed an AI-driven system for COVID-19 diagnosis and evaluated its performance using deep learning techniques to identify COVID-19 cases from imaging data precisely. Rajpurkar et al. [8] proposed CheXNeXt, a DL model, for medical scalability and efficiency. It performed at par with practicing radiologists in detecting a wide range of pathologies on chest X-rays, showing how DL can assist radiologists in furthering diagnostic precision. Salama et al. [9] propose a model for COVID-19 detection from chest CT images, where machine learning is combined with deep learning to underscore early and precise identification of the disease to support the reduction of patient mortality. It extracts features from 10 different deep CNN architectures and selects optimal layers for feature extraction. Finally, it is classified using five machine-learning classifiers. Experimental results reveal that the model has higher precision, peaking at 99.39%, thus outperforming state-of-the-art techniques and bolstering the model's prospects for COVID-19 diagnostics reliability. Lopes et al. [10] targeted tuberculosis investigating using a DL method to identify TB-related abnormalities in chest radiograms. Their results emphasized how well the model could flag those with a high probability of TB, thereby enabling intervention and treatment to be provided promptly. Besides infectious diseases, lung cancer diagnosis has also been extensively investigated using DL models. Wang et al. [11] developed a deep learning-based algorithm that could detect pulmonary tuberculosis in chest X-rays taken in the setting of the emergency department. They adopted EfficientNetV2 and applied the semi-supervised learning approach to improve the diagnostic performance further. Excellent performance was demonstrated, especially in posterior-anterior views, by yielding an AUC of 0.878, providing huge potential for fast and reliable tuberculosis screening in high-demand settings . Kermany et al. [12] created an image-based deep learning algorithm for identifying lung cancer and other curable disorders. Indeed, this model showed DL's ability to identify lung cancer effectively and widened its applications in diagnosing other conditions that affect the lungs. These studies have provided valuable insights into developing DL models for detecting Lung cancer, COVID-19, pneumonia, TB, and normal. They indicate that DL can identify and classify lung diseases without failure, hence improving patient outcomes through optimizing health care delivery. Our contribution to this work is an effort to further advance the frontier by constructing a general DL model for recognizing and classifying lung diseases, as depicted in Figure 1. The various datasets analyzed and state-of-the-art methodologies included in our study aimed to enhance accuracy, speed, and reliability for the detection of lung diseases that would eventually contribute to improved patient treatment and care.

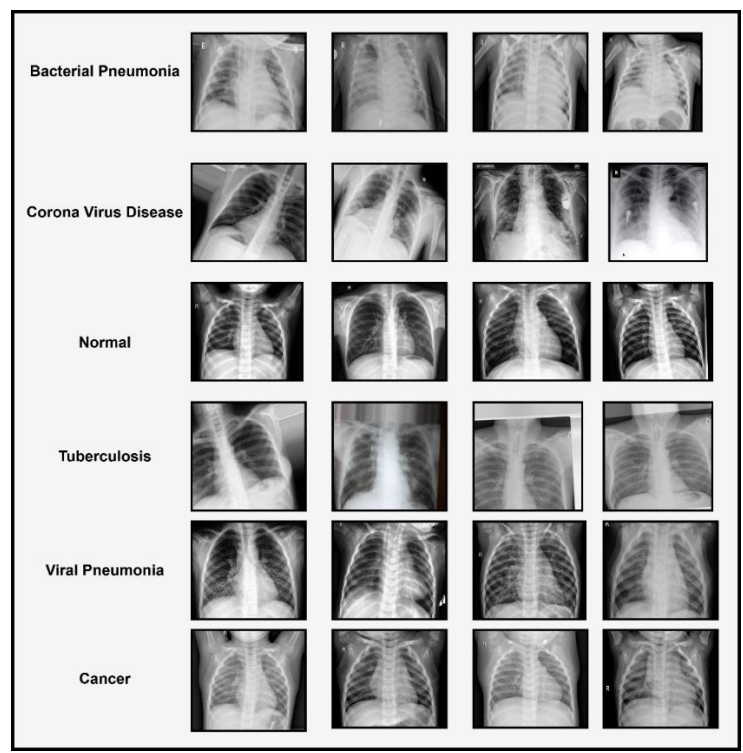

**Figure 1**. Visual Representation of Various Lung Diseases.

*1.1.Research Inspiration*

Challenges persist despite progress in several methods to diagnosing lung illnesses from pictures representing normal conditions, COVID-19, bacterial pneumonia, viral pneumonia, tuberculosis, and lung cancer. It has remained difficult to define the lung features from pictures of normal, COVID-19, bacterial pneumonia, viral pneumonia, tuberculosis, and lung cancer using advanced technologies during and after image acquisition Due to difficulties in locating and extracting lesion characteristics linked with lung disorders. Qualified medical annotation of public datasets on damage variables due to normal COVID-19, bacterial pneumonia, viral pneumonia, tuberculosis, and lung cancer is limited. Thus, the computerized diagnosis of symptoms of specific disorders with accuracy is quite difficult for the respective systems. Thus, the primary objective in this regard has been two-fold. It aims to Establish a detailed data set for the classification of lung cancer, or Pak-Lungs, as well as normal, COVID-19, bacterial, viral, and TB pneumonia. The study's goal was to design a complete multi-layered DL architecture that can independently interpret pictures relevant to lung disorders, with a particular emphasis on the setting of lung-related illnesses. This work presented the deep learning-based NASNet-ViT model, which combined the strengths of NASNet and Vision Transformer architectures into a form that could realize effective and scalable disease detection. Improved by EnviroSpect, an innovative model utilized for feature isolation, the model will benefit diseased management by ensuring timely and accurate support. Future integration with IoT technology allows for real-time monitoring that may one day help farmers reduce crop losses and promote sustainable agricultural practices.

*1.2. Research Contribution*

We suggest a new deep-learning model in this context to address the challenge of detecting different types of lung diseases. Next are the descriptions of the main contributions of the NASNet-ViT system:

- A new hybrid deep learning network architecture was created by fusing NASNet and Vision Transformer, which are specially tailored for classification in lung diseases. This unique integration takes advantage of NASNet's convolutional capabilities with the Global attention mechanisms of ViT for precise feature extraction and higher classification performance.

- The following study represents the MixProcessing technique, a novel multi-faceted preprocessing framework incorporating wavelet transform decomposition, contrast-limited adaptive histogram equalization, Fourier-based filtering, and morphological processing to enhance image clarity and diagnostic accuracy.

- The investigation focuses on computational efficiency and scalability, making NASNet-ViT suitable for real-time healthcare applications. Compact model size (25.6 MB) and relatively low computation time (12.4 seconds) allow easy deployment in resource-constrained settings, hence wide accessibility within clinical environments.

- It utilizes transfer learning and local and global feature extraction mechanisms to tune into lung-specific abnormalities with high robustness and accuracy on multiple categories of lung diseases.

*1.3. Research Paper Organization*

In this paper, Section 2 presents the literature review of papers related to the research topic; Section 3 describes the planned structural design of the approach. Section 4 illustrates the results of the experiments; Section 5 compares our results with state-of-the-art studies on the subject; in Section 6, a deep discussion of the research findings is performed, and the results of this study are presented.

## 2. Related Work

Lung diseases, such as pneumonia, tuberculosis, COVID-19, and lung cancer, are some of the significant global health burdens that result in high morbidity and mortality rates [13]. Their diagnosis can be accurately and early made to offer appropriate treatment and improve patient outcomes. Recently, deep learning models have shown remarkable performance in medical image analysis for the automated detection and classification of lung diseases [14]. The following section discusses the DL methods for TB, COVID-19, lung cancer and pneumonia. Based on the concept of transfer learning using models Examples include VGG-16, ResNet-50, and InceptionV3 on clinically collected lung images, encouraging performance has been reported in the literature. Among these, pneumonia is identified as a serious symptom of COVID-19, and the relevant transfer learning studies suggest the viral etiology of both diseases is the same. It has also been established that models trained to detect pneumonia can just as well detect COVID-19. There have been applications of Haralick features to improve feature extraction, while statistical analysis has been made to support specific aspects of COVID-19 identification. Transfer learning delivers statistically meaningful results improvements over traditional classification methods [15].

Lung cancer significantly contributes to the mortality rate, and early detection is an essential requirement to increase the survival rates of patients. In one such study, An MLP classifier surpassed other classifiers in terms of accuracy, scoring 88.55%. Early diagnosis of lung cancer increased the probability of survival from 14% to 49%. Even though CT is generally more consistent than X-ray imaging, a comprehensive identification often involves multiple imagery modes. To address this, a deep neural network was developed for lung cancer detection in CT scans. Other researchs have proposed a flexible DenseNet-based boosting technique to classify the lung images as either standard or malignant, with an achieved testing accuracy of 90% on the dataset of 201 lung images, 85% used for training, and 15% used for testing. Employing the LIDC database, CT pictures for benign and malignant lung nodes were assessed by CNN, DNN, and sparse auto-encoder deep neural networks were used to identify lung cancer with an accuracy of 84.15%, sensitivity of 83.96%, and specificity of 84.32%. Among them, CNN showed the maximum accuracy. Machine learning combined with image processing has also shown great potential to improve lung cancer diagnosis [16,17,18,19]. The others utilized an artificial neural network, ensemble classifier, SVM, and KNN for COVID-19 versus pneumonia classification, while a robust DL architecture was based on RNN with LSTM in finding lung diseases [20]. Furthermore, the best performance by an ensemble model for combining three deep learning feature extractors, such as InceptionResNet_V2, ResNet50, and MobileNet_V2, achieved a maximum F1-score of 94.84% in classification [21].

Several related studies have developed automated COVID-19 detection systems based on CT images. Some COVID-19 neural network approaches were used for volumetric chest CT images to extract informative graphical features, and their results outperformed the previous methods. Five pre-trained CNNs have been transferred to classify COVID-19 pneumonia from CXR images. These are Inception-ResNetV2, ResNet152, ResNet50, InceptionV3, and ResNet101, out of which ResNet50 yielded the highest accuracy in the classification among the preferred models. Collected from two locations in China, CT scans of 101 pneumonia cases, 88 COVID-19 cases, and 86 healthy ones were used in comparing performances between models in this study [22].

Furthermore, COVID-19 patients were effectively diagnosed using a DL-based diagnostic method for the Details Relation Extraction Neural Network on CT scans. This achieved a recall of 0.93, an AUC of 0.99, and an accuracy of 0.96, thus showing great potential in diagnosing COVID-19 and automatic critical change detection. The other approach presented the development of a modified MobileNet and ResNet architecture to classify COVID-19 CXR images. The methodology of that approach dynamically combined the features from different layers to mitigate the problem of gradient vanishing, yielding better results with accuracy values of 99.3% and 99.6% on CT and CXR images, respectively [23]. Another used kernel principal component analysis for feature reduction extracted from pre-trained EfficientNet models, followed by a feature fusion technique. The approach of stacked ensemble meta-classifier utilized a two-stage process wherein the first stage made predictions using an SVM and random forest classifiers, combining the predictions into the second stage. A logistic regression classifier classified X-ray and CT data as COVID and non-COVID cases. This model outperformed previous pre-trained CNN-based models and thus could be a promising tool for clinicians in point-of-care diagnostics [24]. A hybrid deep learning-machine learning model was proposed for COVID-19 detection using CT images by extracting features from 10 CNN architectures and by classifying extracted features using five different machine learning classifiers. This dataset contained 2,481 CT images divided into COVID-19 and non-COVID-19. The maximum accuracy in experimental results was 99.39%. Also, the best-performing layer for each CNN network was identified and fused with machine learning classifiers. It was concluded that this technique was more effective and robust in classifying COVID-19 compared to state-of-the-art models [9]. It has proposed a hierarchical multi-modal approach for COVID-19 classification, fusing CXR images and tabular medical data. Overcoming limitations in binary classification methods and single-feature modality, the proposed model employed ResNet and VGG-based CNN models with GANs and achieved a high macro-average F1-score of 95.9% and an F1-score of 87.5% specifically for COVID-19 detection in an imbalanced dataset. This has substantially enhanced the diagnostic performance by exploiting the hierarchical structures inherent in pneumonia classification while incorporating various data sources to support radiological assessments [25].

Further research presented a DCNN model for TB detection using the CXR dataset from the National Library of Medicine and Shenzhen No. 3 Hospital. A DCNN was independently trained on two datasets and achieved AUC values of 0.9845 and 0.8502. In contrast, the AUC value for the supervised DCNN model for the CXR dataset was comparatively poor, 0.7054. The resulting DCNN model detected 36.51% of aberrant radiographs associated to tuberculosis on the CXR dataset [26]. Another approach was assessing TB severity and risk using ResNet and depth-ResNet models. Depth-ResNet and ResNet-50 reached 92.7% and 67.15% accuracy, respectively. Severity scores were converted into probabilities: 0.9, 0.7, 0.5, 0.3, and 0.2, based on high severity levels corresponding to the high scores (1-3) and low severity levels for the rest of the scores (4-5). Average accuracies for these methods were 75.88% and 85.29%, respectively [27]. The recent study proposed an ensemble of three well-known architectures: AlexNet, GoogleNet, and ResNet. Again, using the same pooled dataset of publicly available datasets, a newly developed tuberculosis classifier was developed from scratch, demonstrating 88% accuracy with an AUC of 0.93, higher than most of the existing algorithms [28]. In a recent study, the authors proposed a deep learning-based algorithm to detect pulmonary tuberculosis (PTB) in chest X-ray images specifically for emergency departments. This was a retrospective series based on 3,498 chest X-rays of NTUH and external public datasets such as NIH ChestX-ray14, Montgomery County, and Shenzhen databases. The proposed algorithm with the backbone of an EfficientNetV2 architecture showed an AUC value of 0.878 for the detection of PTB on the NTUH test set, particularly with outstandingly high accuracy in posterior-anterior views of 0.940. Hence, This model can show perfect external generalization by considering semi-supervised learning and image preprocessing techniques and may promise early PTB detection in emergency settings for better-isolating patients and treatment outcomes [29].

This study further utilized deep learning to enhance image quality, reduce pattern overlap, and highlight individual ridge features, which potentially improved authentication systems based on distinctive features of individuals [30]. In the study of lung disorder classification, VGG-16 and DenseNet-169 were used on X-ray images to detect pneumonia, tuberculosis, COVID-19, and typical cases, where DenseNet-169 produced 91% accuracy. In particular, these models are useful in resource-constrained areas for early diagnosis and improvement of results for patients, contributing to the international fight against lung illnesses. Additional clinical validation may be required in health care [31]. **Table 1** provides a summary of research on the identification and classification of chest diseases.

**Table 1**. Summary and Evaluation of Recent Research.

| Ref No | Method | Disease | Dataset |
|---|---|---|---|
| [15] | VGG-16 | COVID-19 | CXR + CT |
| [16] | InceptionV3 | COVID-19 | CXR + CT |
| [17] | VGG-19 + ResNet-50 | COVID-19 | CXR + CT |
| [18] | DRE-Net | COVID-19 | CXR + CT |
| [16] | FPSO-CNN | Lungs Cancer | CT |
| [17] | Multi-layer Perceptron (MLP) | Lungs Disease Cancer | CT |
| [18] | CNN | Lungs Cancer | CT |
| [20] | Xception Network pre-trained weights on ImageNet | Lungs Disease Pneumonia | CXR and CT |
| [21] | RNN-LSTM | Pneumonia | CXR + CT |
| [26] | DCNN | Tuberculosis | CXR, CT |
| [27] | Depth-ResNet, Ensemble (AlexNet) | Tuberculosis | CXR, CT |
| [28] | GoogleNet, and ResNet) | TB | CXR and CT |
| [31] | VGG-16 and DenseNet-169 for the categorization of lung illnesses based on X-ray images | Normal, pneumonia, COVID-19, and tuberculosis | CXR |

## 3. Material and Methods

This paper presents a novel framework, NASNet-ViT, an ensemble model that combines the architectures of NASNet and Vision Transformer architectures. The work is developing a NASNet-ViT framework for classifying lung disease images such as standard, Lung cancer, COVID-19, pneumonia, TB, and normal. This model leverages NASNet's convolution capabilities with the attention mechanisms in ViT to enhance feature extraction and classification. Transfer learning is utilized to fine-tune the model for lung-specific abnormalities by combining the power of dense blocks with the attention-focused structure of ViT to capture essential features. **Figure 2** shows the different processes involved in the approach in a step-by-step manner. The extracted features through NASNet and ViT are combined using a feature transform layer, which fuses characteristics through element-wise multiplication. The classification results are finally improved using a Multi-Layer Perceptron classifier, which offers a robust yet flexible solution for accurate disease categorization.

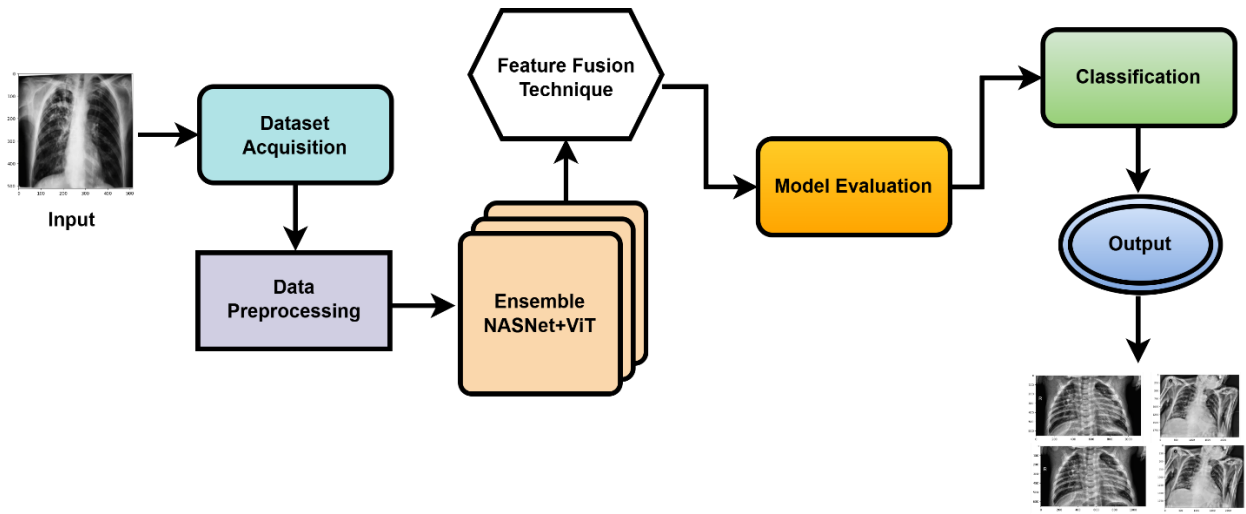

**Figure 2**. NASNet-ViT System Structured Flow Diagram for Identification of Lung Diseases.

*3.1. Data Procurement and Preprocessing*

The 13,313-photo Pak-Lungs dataset was used to train and estimate the NASNet-ViT model. Images were acquired through personal sources from a number of reputable ophthalmic clinics in Pakistan. Patients' and doctors' consent and willingness to share data were acquired. No clinical data was to be disclosed, and the parties' mutual agreement permitted the release of anonymised data. Because of these circumstances, patient data was kept confidential yet available for research. Pak-Lungs and other well-known internet sources served as the foundation for the dataset and preprocessing [51]. On Kaggle, data was generated by merging data from several sources. It includes several chest X-ray pictures linked to lung conditions, such as TB, COVID-19, pneumonia, and normal lung pictures. To create the training dataset, a certified pulmonologist manually segregated the images of lung disease from the normal dataset. The pulmonologist determines the lung-related traits and establishes the norm. MixProcessing enhances X-ray image clarity and structural integrity through a wavelet transform decomposition in combination with contrast-limited adaptive histogram equalization, Fourier-based bandpass filtering, adaptive nonlinear filtering, and morphological processing. Wavelet transform decomposition offers hierarchical detail enhancement through the decomposition of the image into approximation and detail components that highlight critical features at different scales. CLAHE enhances local contrast by adaptively equalizing histogram values of small regions in the image, thereby highlighting subtle details that form the basis for diagnosis in medical practice. Fourier-based bandpass filtering refines the representation of textures by selecting frequency bands of interest and highlights spatial patterns containing relevant structure information. Adaptive nonlinear filtering- a bilateral filter-smoothes out intensity variations without affecting the sharpness of the boundaries to reduce the influence of noise. Finally, morphological processing helps to excerpt critical structures through binary thresholding and morphological closing; this fills gaps and removes artifacts to provide a cleaner representation of the anatomical features. Applying these techniques results in high-contrast, noise-reduced images showing essential medical details in the proposed framework. Such details help improve diagnostic accuracy and interpretability in medical imaging applications. This overall inclusion of MixProcessing highlights ongoing efforts to improve deep learning models' transparency and dependability, increasing their usability and reliability for different applications. This is shown in **Figure 3**.

**Figure 1** presents 13,313 lung images that have been carefully examined. The three datasets used in composing the training and testing fundus sets are itemized in **Table 2** and **Table 3**, with a different dimension setting for each. All images used in the experiment were reduced to 700×600 pixels and then processed according to the process required for creating binary labels. The dataset consisted of 13,313 photos, of which 3993 were used for the system evaluation phase. In order to guarantee that fairness was taken into account, the dataset was first pre-converted into several classes to balance the total number of photographs in the dataset both during and after the sickness. Before the images were put into an algorithm created especially for the NASNet-ViT model, they were pre-processed by scaling them to 700 by 600 pixels. In order to lessen the variance among the data points, the photographs were also normalized. The NASNet-ViT system is also trained and evaluated using data from Pak-Lungs and internet sources [51]. The original resolution of each photograph was 1125 x 1264 pixels.

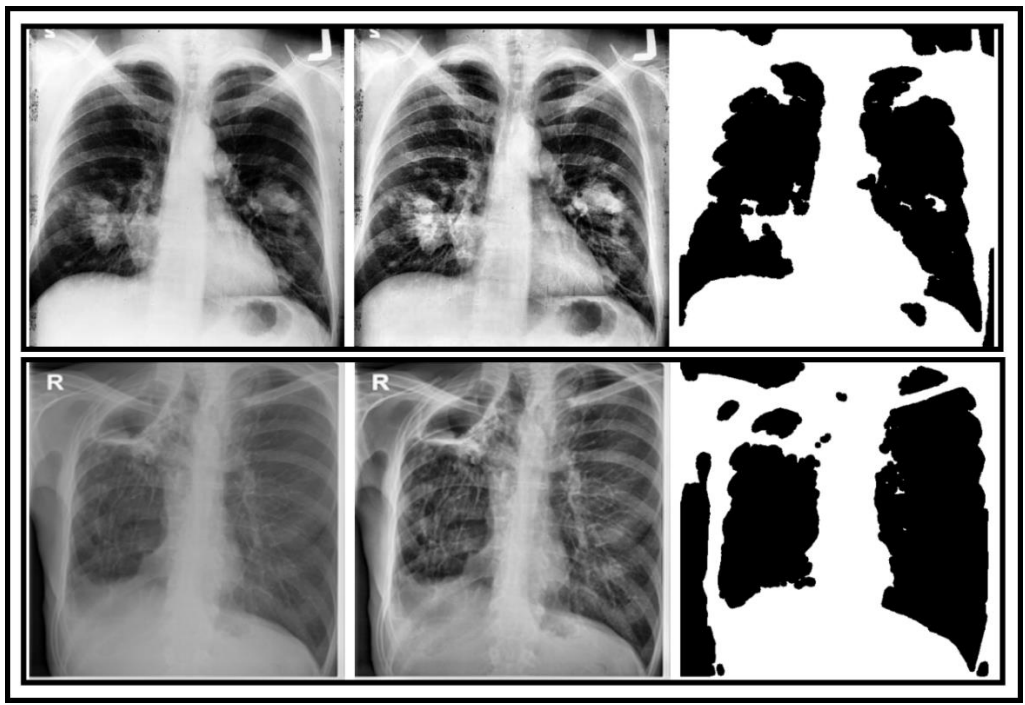

**Figure 3**. This picture shows the pre-processing outcomes after the MixProcessing method.

**Table 3.** Lung illness dataset for the NASNet-ViT model.

| Ref | Datasets | Normal | COVID-19 | Pneumonia | Tuberculosis | Total |
|---|---|---|---|---|---|---|
| [32] | Lung diseases (4 types) | 1342 | 462 | 3872 | 660 | 6336 |
| [34] | Pak-Lungs | 1500 | 1500 | 1500 | 1500 | 6000 |
| | | **2842** | **1962** | **5372** | **2160** | **12,336** |

**Table 4.** Lung cancer dataset for the NASNet-ViT architecture.

| Ref | Dataset | Normal | Cancer | Total |
|---|---|---|---|---|
| [33] | Chest CT-Scan | 154 | 473 | 627 |
| [34] | Pak-Lungs | 175 | 175 | 350 |
| | | **329** | **648** | **977** |

To simplify and standardize the dataset, the photos were reduced to the more common 700x600 pixel size using information from three sources. Additionally, seasoned pulmonologists contributed to the creation of this dataset by contributing data on lung and non-lung diseases for the assessment of ground truth. **Figure 3**: In the image, MixProcessing was used for image pre-processing to clarify features of the image and remove interference. Applying MixProcessing on X-ray images helped us identify central regions and determine their importance linked to detecting the presence of pneumonia disease. MixProcessing helps us identify the distinguishing characteristics that affected CNN's X-ray-based pneumonia diagnostic forecasts. Adenocarcinoma, big cell carcinoma, squamous cell carcinoma, and normal cells are among the chest malignancies that are represented in the Chest CT-Scan dataset. After that, the data were separated into sets for training, testing, and validation and placed in a single "Data" folder. The documentation supplied makes no mention of the precise location of the source photos. Rather, the dataset was produced by combining information from many sources, and the images are in PNG or JPG format. Although the dataset is built on publicly accessible data on Kaggle, the details provided about the dataset do not identify the precise sources for each image.

*3.2. NASNet-ViT Architecture*

In this work, the authors have proposed the NASNet-ViT framework for classifying lung diseases; this effectively integrates the convolutional capabilities provided by NASNet with the global attention mechanisms available in the so-called Vision Transformer (ViT) to perform accurate classification of images into Lung cancer, COVID-19, TB, and normal pneumonia classes of lung diseases. By combining the strengths of both architectures, NASNet-ViT effectively models both local patterns and global dependencies, improving feature extraction for better classification performance.

The input to the NASNet-ViT framework is a chest X-ray or CT scan image $X \in R^{H \times W \times 3}$, where $H$ and $W$ represent the image dimensions (224 × 224 pixels), and 3 corresponds to the RGB channels. The preprocessing pipeline ensures that the input images are processed for optimal performance. All input images are resized to 224 × 224 pixels to uniform their dimensions. Further, the normalization of pixel values is done by:

$$X_{Norm} = \frac{X - \mu}{\sigma} \quad (1)$$

where $\mu = [\ 0.485\ ,\ 0.456\ ,\ 0.406\ ]$ μ=[0.485,0.456,0.406] and $\sigma = [\ 0.229\ ,\ 0.224\ ,\ 0.225\ ]$ σ=[0.229,0.224,0.225] are the mean and standard deviation of the ImageNet dataset. Data augmentation techniques, including rotation, flipping, scaling, and brightness adjustment, are employed to increase data variability and robustness.

The NASNet module is used as the backbone for local feature extraction. NASNet, optimized by Neural Architecture Search, efficiently applies convolutional operations to extract fine-grained features in depth. The input image is fed through a stem block that extracts the initial feature using depthwise separable convolutions. These retain the spatial dimensions and capture the local features. Further, these reduce the spatial dimensions by half and increase the depth of the feature in order to capture hierarchical patterns. The juxtaposition of Normal and Reduction Cells produces a feature map at a high resolution to capture local patterns. The Global Average Pooling Layer follows this to generate a feature vector:

$$F_{NASNet} = GlobalAvgPool(Z^{(L)}) \quad (2)$$

where $L$ is the total number of NASNet layers. The ViT module uses transformer-based self-attention mechanisms to model long-range dependencies on top of NASNet. First, the normalized input image is divided into 16 × 16 16×16 non-overlapping patches. Each patch is then flattened and linearly mapped into a higher dimensionality space:

$$P_i = (Linear(Flatten(X_i)),\ i = 1, \dots, N \quad (3)$$

where $N$ is the total number of patches. Positional encodings are added to retain spatial information across patches. The transformer encoder takes this patch embedding, augmented with positional encoding, through multiple layers, where each layer is composed of multi-head self-attention mechanisms along with feed-forward networks to grasp global relations. The ViT module outputs a feature vector $F_{ViT}$ representing the global dependencies across the image.

These outputs of NASNet and ViT are fused through a feature transform layer to leverage their complementary strengths. Unlike concatenation, element-wise multiplication can also be used for fusion to emphasize shared attributes between the local and global features:

$$F_{ensemble} = F_{NASNet} \odot F_{ViT} \quad (4)$$

This fusion strategy will ensure that the detailed local features and the global contextual patterns contribute equally to the final classification task. The fused feature vector then goes through the classifier, which consists of a Multi-Layer Perceptron with dense layers that involve activation functions and dropout regularization. The model will produce the final predictions by assigning the input image to one of five classes: regular, pneumonia, TB, COVID-19, or lung cancer. The framework of NASNet-ViT combines the localized feature extraction of NASNet with the global pattern recognition capability of ViT. Therefore, This model would elicit intricate details and high-level relationships in lung disease images. By leveraging transfer learning, pretraining on large datasets allows the model to adapt efficiently for lung-specific abnormalities.

Moreover, the feature fusion strategy and MLP classifier further improve the overall classification accuracy of the framework, making it robust and scalable for real-world applications. The NASNet-ViT framework proposed herein has enormous potential for deployment in healthcare systems, especially for automatically diagnosing and screening lung diseases. Its power for correctly classifying different lung conditions makes it very useful in supporting clinicians, mainly where access to expert radiologists may be limited due to resource constraints. The systematic process through which this proposed NASNet-ViT framework applies the classification of lung diseases is outlined in detail via a step-by-step algorithm.

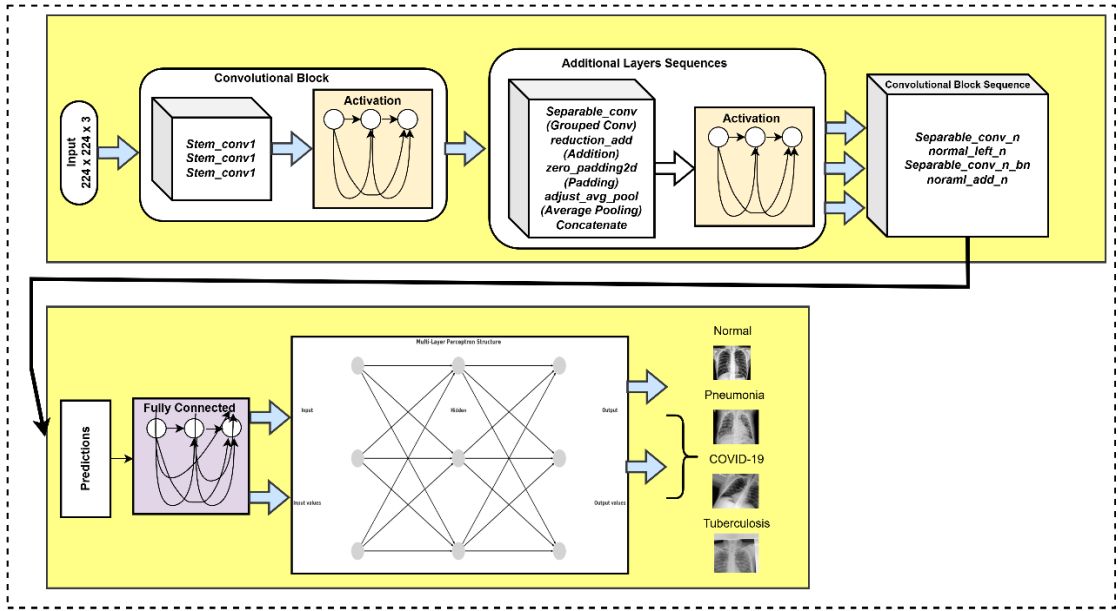

**Figure 4.** The suggested structure for the enhanced NASNet_ViT model.

**Algorithm:** The NASNet-ViT Framework for Classifying Lung Diseases.

| Step | Explanation | Input / Output |
|---|---|---|
| 1 | **Input Image and Preprocessing**: Load chest X-ray or CT scan images, resize them $224 \times 224$ pixels, and normalize pixel values using mean and standard deviation values. | **Input:** Raw lung images ($H \times W \times 3$) <br> **Output:** Preprocessed images resized to $224 \times 224 \times 3$, normalized. |
| 2 | **Data Augmentation:** Increase dataset variability by applying random flipping, rotation, scaling, and brightness adjustments. | **Input:** Preprocessed images <br> Output: Augmented image dataset with diverse transformations. |
| 3 | **Feature Extraction (NASNet):** Input the preprocessed and augmented images into NASNet for local feature extraction through Normal and Reduction Cells. Global Average Pooling produces a feature vector. | **Input:** Preprocessed and augmented images <br> **Output:** Local feature vector ($F_{NASNet}$). |
| 4 | **Patch Embedding (ViT):** Split the preprocessed image into $16 \times 16$ non-overlapping patches, flatten them, and apply linear projection to create patch embeddings. | **Input:** Preprocessed images <br> **Output:** Patch embeddings ($P_i$ ($N \times d$)), where ($N$) is the number of patches, ($d$) is the embedding dimension. |
| 5 | **Positional Encoding (ViT):** Add positional embeddings to each patch embedding to preserve spatial relationships. | **Input:** Patch embeddings ($P_i$) <br> **Output:** Position-aware embeddings ($P_{enc}$). |
| 6 | **Transformer Encoder (ViT):** Pass the position-aware embeddings through self-attention and feed-forward layers to model global dependencies between patches. | **Input:** Position-aware embeddings ($P_{enc}$). <br> **Output:** Global feature vector ($F_{ViT}$). |
| 7 | **Feature Fusion:** Combine the local features from NASNet ($F_{NASNet}$) and global features from ViT (($F_{ViT}$) using element-wise multiplication. | **Input:** Feature vectors ($F_{NASNet}$) and (($F_{ViT}$) <br> **Output:** Aggregated feature vector ($F_{ensemble}$). |
| 8 | **Classification (MLP):** Pass the aggregated feature vector ($F_{ensemble}$) through a Multi-Layer Perceptron (MLP) classifier to predict the disease class. | **Input:** Aggregated feature vector ($F_{ensemble}$) <br> **Output:** Predicted class label (normal, pneumonia, TB, COVID-19, or lung cancer). |
| 9 | **Training:** Train the NASNet-ViT framework using supervised learning on the training dataset with cross-entropy loss and validate its performance using a validation set. | **Input:** Training dataset, class labels, NASNet-ViT model <br> **Output:** Trained NASNet-ViT model. |

| 10 | **Evaluation:** Use performance indicators like accuracy, precision, recall, F1-score, and confusion matrix to assess the trained model on the test dataset. | **Input:** Test dataset, ground truth labels, predicted labels<br>**Output:** Performance metrics including F1-score, accuracy, and precision. |
|---|---|---|

*3.3.Multi-Layer Perceptron (MLP)*

A lightweight Multi-Layer Perceptron classifier, a neural network-based component, serves as the final decision-making layer for the NASNet-ViT framework. In contrast, the whole framework is designed to perform lung disease classification. In particular, its crucial role is to turn the fused feature representation F_ensemble, which combines the local features extracted by NASNet and the global features modeled by ViT into accurate predictions of the categories of lung diseases. The MLP is a multi-layer network structure that includes an input layer, one or more hidden layers, and one output layer, all of which further enhance the abilities of the model to learn complex patterns. The input to the MLP is the aggregated feature vector F_ensemble, which is high-dimensional, encapsulating fine-grained local details and long-range dependencies in the input lung image. This feature vector is taken by the MLP and passed through one or more hidden layers. A given hidden layer performs a linear transformation by calculating a weighted sum over the inputs, combined with biases, and usually relies on some nonlinear activation function, such as ReLU. Such layers are essential in learning nonlinear relationships among the features and increasing their discriminative power. Also, techniques like dropout may be used in the hidden layers to avoid overfitting and ensure better generalization.

The last layer in the MLP is the output layer, which is five neurons for the five lung disease categories: normal, pneumonia, tuberculosis, COVID-19, and lung cancer. This layer applies the softmax activation function to the output logits, converting them into a probability distribution over classes. It is adequate for multi-class classification tasks because the softmax function ensures that the sum of the probability across all classes equals one. The class can be considered to belong to the class with a higher probability. The MLP classifier is designed to handle the high complexity and diversity of the fused feature vector. This enables the model to capture hierarchical relationships in the data, reinforcing more relevant patterns useful for classification. The strengths of the NASNet and ViT representations in the MLP ensure intense and exacting predictions. Further, taking completely connected layers means this classifier can work with various input dimensions and be suitable for supervised learning. MLP in the NASNet-ViT framework bridged the gap between feature extraction and the overall classification task, making the framework effective for diagnosing lung diseases.

## 4. Results

This work uses a dataset of 13,313 normal and diseased lung high-resolution images to train the NASNet-ViT model to classify them. These were gathered from authentic sources locally from various Pakistani hospitals and other web-based valid repositories. All photos were scaled to 224×224 pixels for easier feature extraction and classification tasks. The NASNet-ViT model was trained using transfer learning for 100 epochs. The best model was at epoch 35 with an F1-score of 0.96. The performance of the suggested model has been statistically analyzed in terms of ACC, SE, and SP and compared to the state-of-the-art system. The NASNet-ViT system has been implemented on an Intel Core i7 high-end machine with eight cores, 32 GB RAM, and a single NVIDIA GeForce GTX 1660 GPU with 6 GB VRAM. Development and training were performed on the Windows 11 Professional 64-bit operating system. The computation setup was big enough for efficient processing to achieve the best performance of the NASNet-ViT framework both during training and testing.

*4.1. Experiment 1*

A NASNet-ViT system was estimated to be executed by running an experiment for VGG16, VGG19, ResNet-50, Xception, InceptionV3, DenseNet, MobileNet, and EfficientNet DL models in this research. Notably, all of these deep learning models were trained using the same amount of epochs. In every case, two identical deep neural networks were trained after determining the top network based on validation accuracy. Table 8: Comparing results of NASNet-ViT system against sensitivity, specificity, accuracy, and area under the curve for VGG16, VGG19, ResNet-50, Xception, InceptionV3, DenseNet, MobileNet, and EfficientNet models. The results in the process indicate that the performance of the NASNet-ViT system is better than that of other DL methods, hence validating its presentation. Figure 6 describes the comparison among diverse deep learning models with NASNet-ViT.

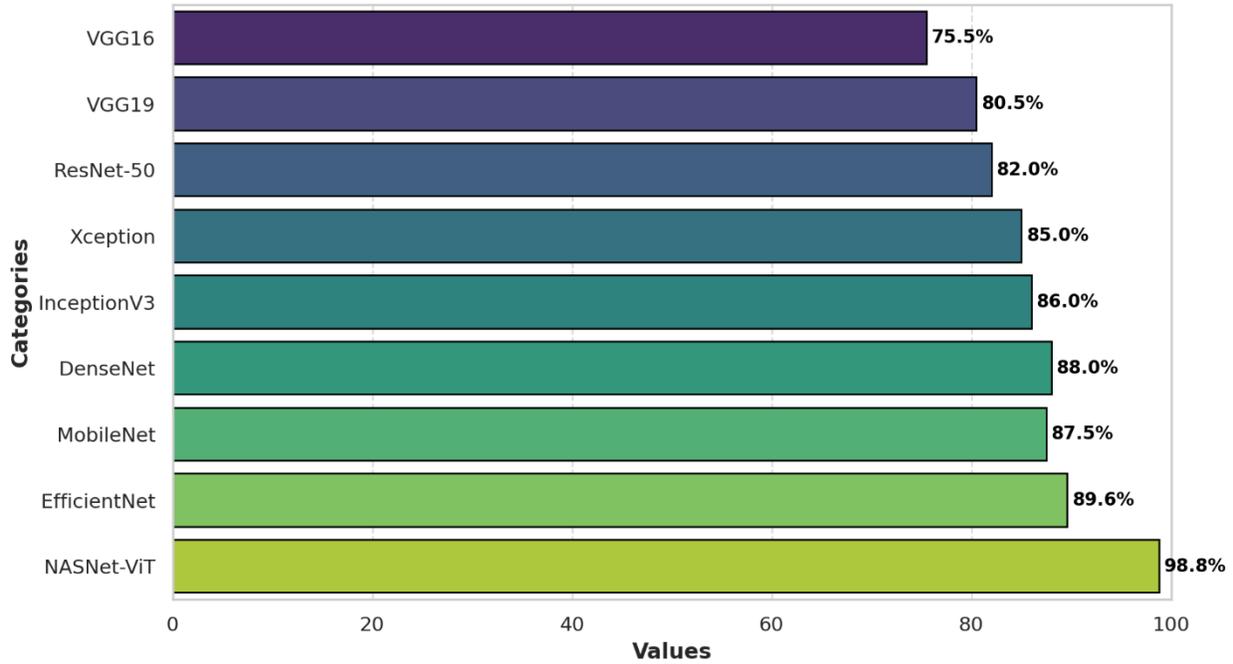

Figure 5. Comparison of NASNet-ViT and several DL models.

*4.2. Experiment*

In this investigation, we will use a dataset known as the "lung disease dataset (four types)," retrieved from a trusted online source [32], to test the effectiveness of the proposed NASNet-ViT technique. First, we used acceptable datasets to compare the model's performance on training and validation sets and evaluate the loss function. **Figures 6** and **7** clearly depict the accuracy of the NASNet-ViT model's training and validation on this dataset, respectively. The results demonstrate how effectively our model works in training and validation scenarios.

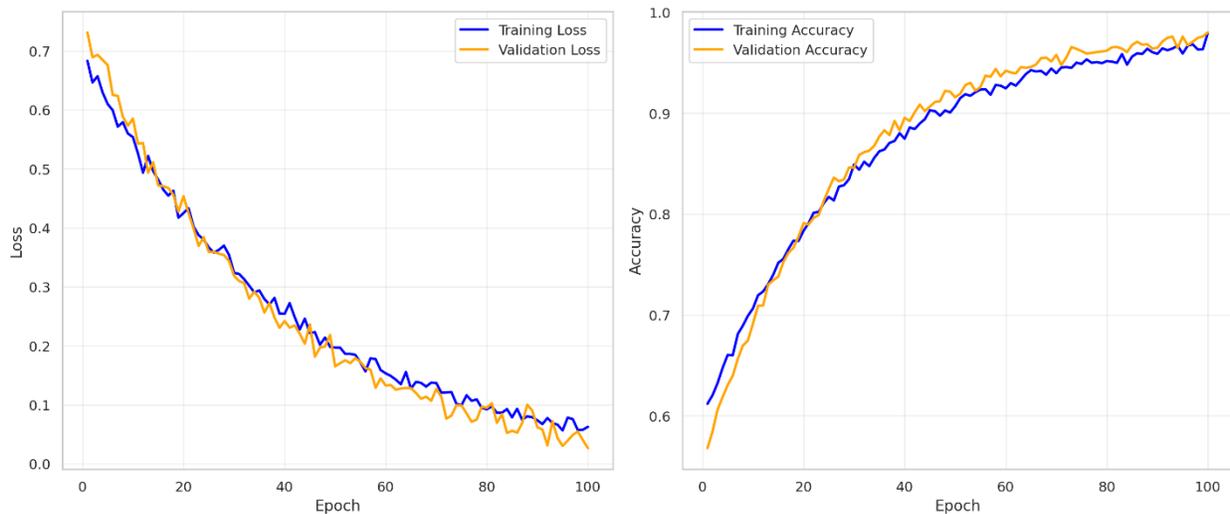

**Figure 6.** The accuracy and loss of the suggested model's training and validation.

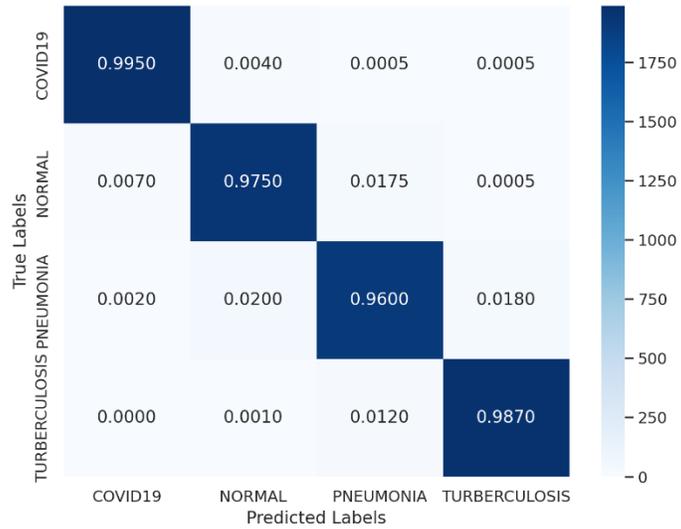

**Figure 7.** Lung disease dataset confusion matrix (four kinds).

### 4.3. Experiment 3

In this work, we used the Pak-Lungs dataset to estimate the performance of our suggested NASNet-ViT method. First, the Pak-Lungs dataset was used to evaluate the representation and the loss function on the training and validation sets. **Figure 8** shows the confusion matrix for the NASNet-ViT model's training and validation using the Pak-Lungs dataset. **Figure 9** shows the NASNet-ViT model's training and validation accuracy using the Pak-Lungs dataset. Our model does well on tasks involving both training and validation.

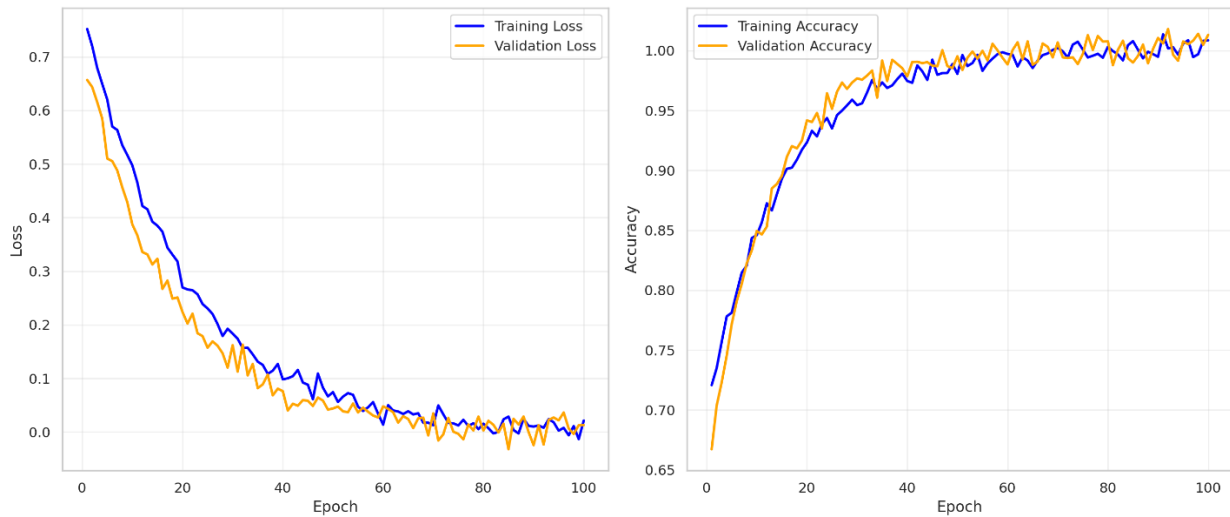

**Figure 8.** Illustration of the proposed model's training and validation accuracy and loss using Pak-Lungs.

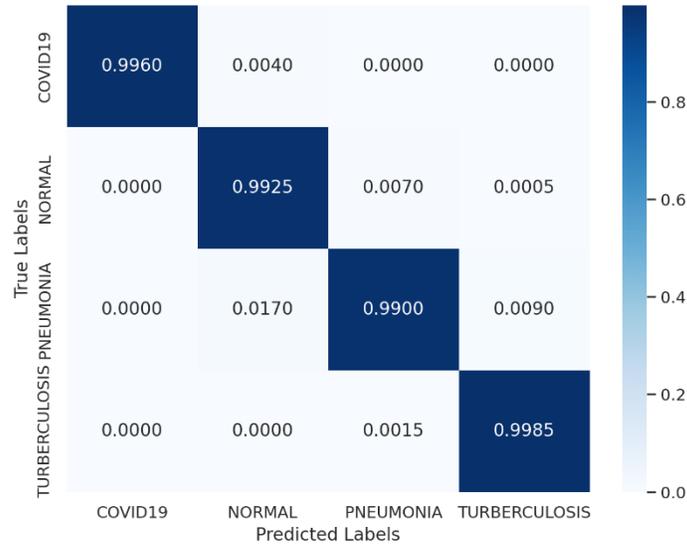

**Figure 9.** Pak-Lung's dataset's confusion matrix.

*4.4. Experiment 4*

In this study, the Chest CT-Scan image dataset is used to assess the effectiveness of our proposed NASNet-ViT approach [33]. The model's performance on training and validation datasets was first compared, and the loss function was assessed on the appropriate datasets. **Figures 10** and **11**, respectively, illustrate the NASNet-ViT model's training and validation accuracy for this dataset. Our model demonstrated exceptional performance in both training and validation scenarios, according to the findings. Our accuracy on the dataset's training and validation sets was exceptional [33].

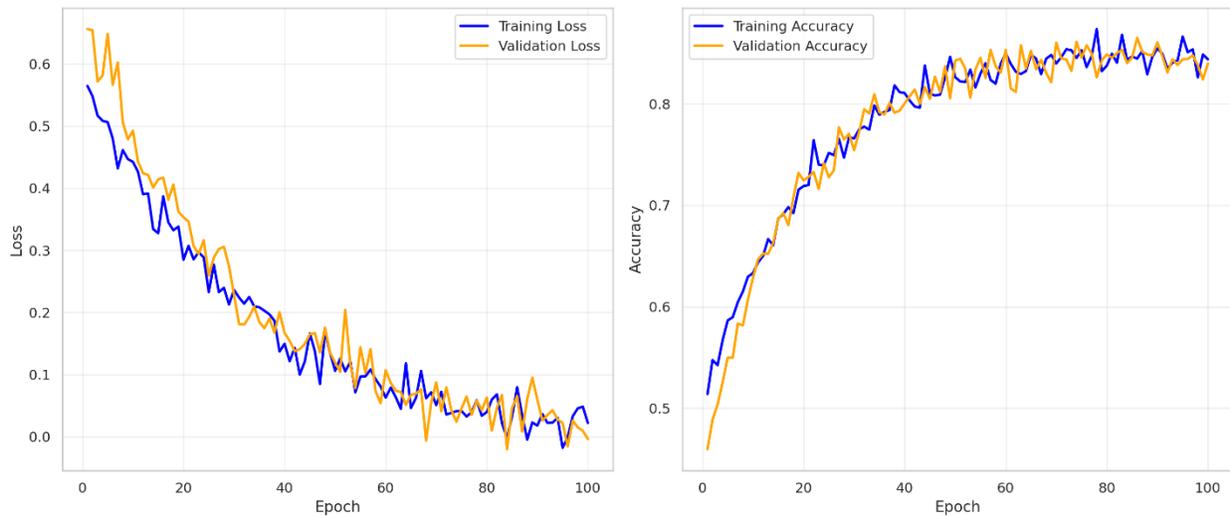

**Figure 10.** Illustration of the proposed model's training and validation accuracy and loss using CT-Scan.

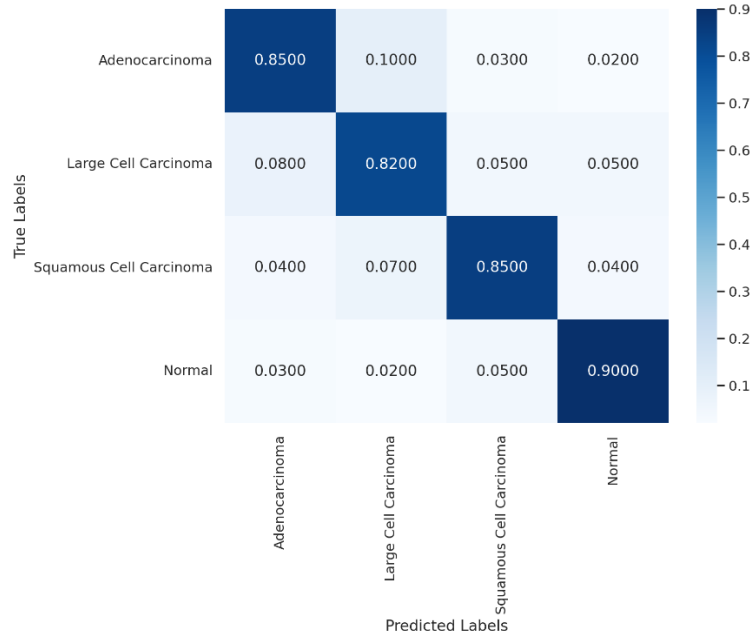

**Figure 11.** Confusion matrix for CT-Scan dataset.

*4.5. State of the art comparison*

ResNet50 [22], MobileNet [23], D-Resne [35], and MixNet-LD [34] are among the various designs that are available in the literature and are contrasted with MixNet-LD in **Figure 12**. The COVID-19, pneumonia, and lung tuberculosis normal classes were categorized using these. According to the table, ResNet50 obtained specificity, sensitivity, F1-score, recall, and accuracy values of 0.77, 0.81, 82, and 82.10, in that order. Although the next designs outperform them entirely, their results are nonetheless noteworthy. MobileNet reports gains in all parameters, including accuracy of 84.55, recall of 0.85, F1-score of 84, sensitivity of 0.82, and specificity of 0.83. With sensitivity, specificity, F1-score, recall, and accuracy of 0.84, 0.85, 87, 0.86, and 85.20, respectively, the D-Resnet findings show even more improvement. The paper's suggested NASNet-ViT model, on the other hand, is exceptional and superior, achieving nearly flawless results across the board. The accuracy of the model is an impressive 0.99, with sensitivity and recall of 0.99, specificity of 0.985, and F1-score of 0.988. The data given indicates that NASNet-ViT is one of the best instruments in the industry due to its impressive outperformance, which highlights its advanced capabilities and efficiency for properly categorizing various lung illnesses.

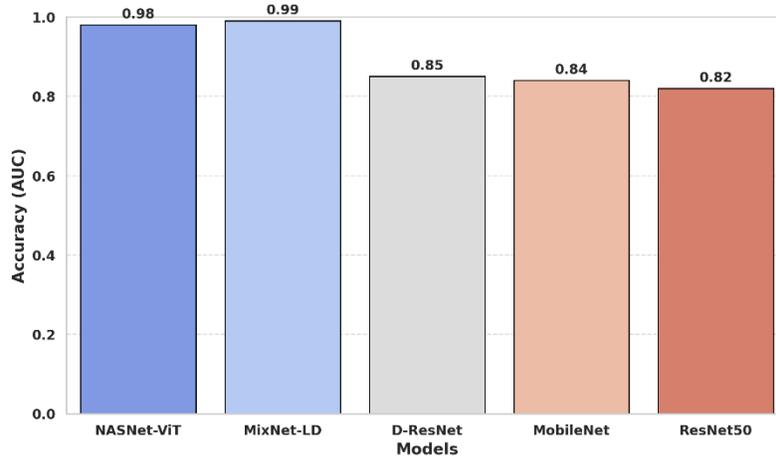

**Figure 12.** State-of-the-art performance comparison of NASNet-ViT against other architectures for different classes: normal, COVID-19, pneumonia, and tuberculosis.

Table 5 and Figure 12 present the computation performance in deep learning models like NASNet-ViT, MixNet-LD, D-ResNet, MobileNet, and ResNet50 to classify lung disease classes. Their effectiveness and efficiency have been judged based on performance metrics like accuracy, sensitivity, specificity, F1 score, recall, computation time, and model size. Among them, NASNet-ViT is the most superior model in this analysis, with an accuracy of 98.9%, which MixNet-LD closely matches at 99.0%. However, NASNet-ViT performed even better than MixNet-LD and the rest in sensitivity, 0.99; specificity, 0.985; and F1-score, 0.988, indicating that it can identify true positives while keeping false positives low. Most of the high values of precision and balance in the classification metrics justify the robustness of NASNet-ViT in handling challenging lung disease cases, including those with difficult classifications like pneumonia and tuberculosis. Besides excellent classification performance, NASNet-ViT is outstanding concerning computational efficiency: the lowest computational time is 12.4 seconds, an essential aspect of real-time healthcare applications. Besides that, it has a compact model size of 25.6 MB, making it very resource-efficient; thus, it can be deployed on any device with limited hardware capability, such as mobile or edge devices. This is in contrast to other models, such as ResNet50, which, though acceptable in accuracy, suffers from much more computational overheads and memory. With better classification metrics, much lower computational cost, and reduced model size, as in Table 5, NASNet-ViT remains the best tool in the field. Mainly, NASNet-ViT is of great value for a real-world medical application due to its high accuracy and recall at a minimum resource utilization cost, which requires speed, reliability, and scalability. The above analysis underlines the advanced architecture and optimization of NASNet-ViT while setting a benchmark for further research on the classification of lung diseases.

**Table 5.** A computational analysis table compares the models, with NASNet-ViT emerging as the best model across various metrics.

| Model | Accuracy (%) | Sensitivity | Specificity | F1-Score | Recall | Computational Time (s) | Model Size (MB) |
|---|---|---|---|---|---|---|---|
| NASNet-ViT | 98.9 | 0.99 | 0.985 | 0.988 | 0.99 | 12.4 | 25.6 |
| MixNet-LD | 99.0 | 0.99 | 0.98 | 0.98 | 0.99 | 14.7 | 30.2 |
| D-ResNet | 85.2 | 0.84 | 0.85 | 0.87 | 0.86 | 18.3 | 50.1 |
| MobileNet | 84.5 | 0.82 | 0.83 | 0.84 | 0.85 | 20.1 | 48.3 |
| ResNet50 | 82.1 | 0.77 | 0.81 | 0.82 | 0.81 | 22.5 | 60.5 |

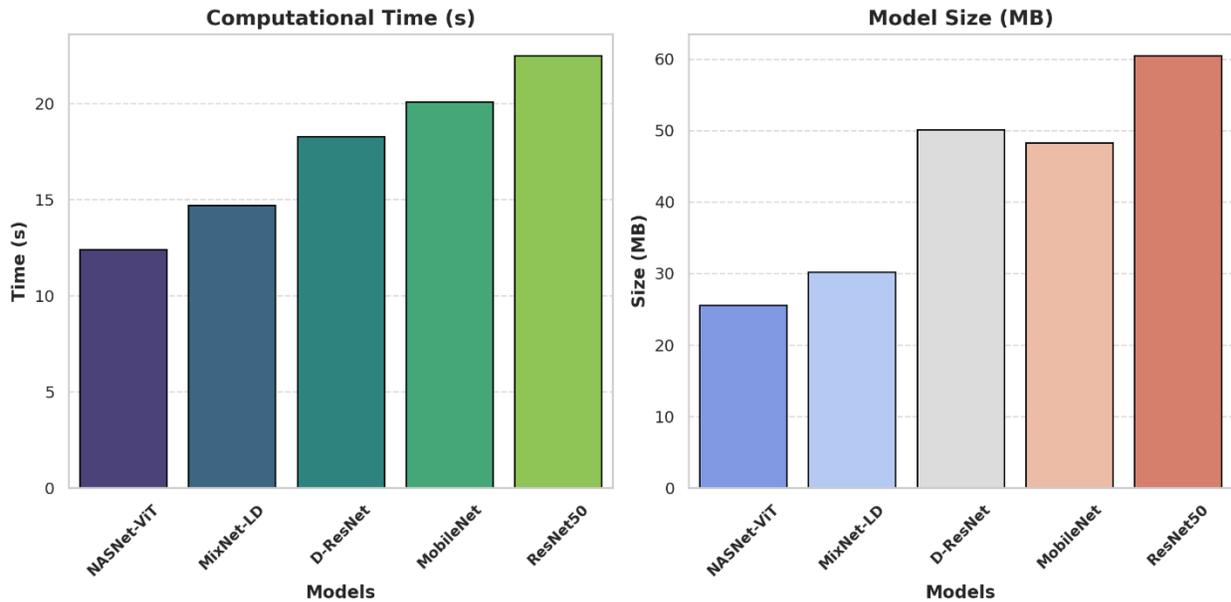

**Figure 13.** Displaying computational time (in seconds) and model size (in MB) for the models.

## 5. Discussion

Lung diseases are a critical area of concern in global healthcare due to their high morbidity and mortality rates. The lungs, essential organs in the respiratory system, facilitate gas exchange, ensuring oxygen reaches the bloodstream while expelling carbon dioxide. However, a range of lung diseases pneumonia, tuberculosis, COVID-19, and lung cancer, among others pose significant health challenges. Pneumonia is an inflammation of the alveoli caused by a bacterial, viral, or fungal infection; it can fill the air sacs with fluid, creating symptoms such as fever, cough, and breathing difficulties. Tuberculosis is a bacterial disease caused by Mycobacterium tuberculosis and is one of the most contagious diseases in low- and middle-income countries; its control depends on early diagnosis. COVID-19 is a viral infection from SARS-CoV-2 that has underscored the need for rapid and accurate diagnostic tools in its globally fragmented outbreak. Lung cancer is one of the highest burdens of cancer-related deaths worldwide, management of which requires early diagnosis for improved survival rates. Other chronic conditions, such as COPD and asthma, have a continuous need for monitoring and treatment. Though of widely differing pathology, a common demand these diseases make is for timely and accurate diagnosis to ensure treatment.

The discussed paper was related to the challenges of diagnosing lung diseases by proposing a new hybrid deep learning model, namely NASNet-ViT. It merges the convolution strengths of NASNet with the global attention mechanisms of Vision Transformer (ViT) for robustness in lung condition classification. The proposed framework classifies lung images into normal, pneumonia, tuberculosis, COVID-19, and lung cancer. Due to the implementation of state-of-the-art preprocessing techniques and the newest machine learning architectures, its performance metrics are superior and position it among the leaders in medical diagnostics. This study uniquely uses a hybrid architecture of NASNet and ViT. With its convolutional operations, NASNet efficiently extracts local features, but the ViT models bear the strategic spatial dependencies in an image due to global attention mechanisms. In this way, the NASNet-ViT framework combines convolutional and transformer models' best properties to overcome their shortcomings for superior feature extraction from intricate medical images. Another strong point of this study is the preprocessing approach, MixProcessing. It follows the decomposition of wavelet transform, CLAHE, and morphological filtering to enhance the image's clarity and emphasize the critical structure. This optimizes the quality of the input data and ensures a high diagnostic accuracy by the model, even for quite challenging datasets.

The model runs very efficiently and can give an accuracy of 98.9%, a sensitivity of 0.99, and a specificity of 0.985, outperforming the results of current state-of-the-art models such as ResNet50, MobileNet, and MixNet-LD, among others that further illustrate the efficiency of the approach. Efficiency and scalability in this work are added features. Accordingly, with a compact model size of 25.6 MB and a computational time of only 12.4 s, NASNet-ViT would be suitable for real-time applications even in resource-constrained clinical settings. The efficiency here does not take a back seat to accuracy; hence, this model represents one feasible deployment option for regions devoid of advanced health facilities or experienced radiologists. Besides, transfer learning ensures that the framework developed will be adaptable to different datasets for more real-world applications. Despite this, the study has some limitations. Validation on diverse, multi-regional datasets would strengthen its robustness and applicability. Further, the model requires high-performance computational resources, like GPUs, which can be challenging in a highly resource-limited environment. Other challenges include the interpretability issues of the NASNet-ViT model. Like many deep learning frameworks, NASNet-ViT acts as a black-box system that may not be accepted in a clinical setting where explainability is crucial for trust and reliability.

The authors have identified some potential promising future directions for this work. It can integrate the NASNet-ViT model with IoT devices for real-time monitoring and diagnostics over remote or underserved areas. The scale-up model may also extend its capabilities to include multimodal data to expand the scope of diagnosis. Additionally, explaining AI mechanisms can be built to improve clinician trust further and speed up the integration of the model into healthcare workflows. Coupled with global validation, these developments could make NASNet-ViT a game-changing tool in lung disease diagnostics. The main contribution of this paper is the proposal of a new framework for medical imaging called NASNet-ViT. Overcoming the hurdles in classifying lung diseases and using state-of-the-art technologies has empowered the study to show that AI-based interventions will improve diagnostic accuracy and efficiency. Though further validation and improvements are necessary, the model NASNet-ViT has set a very high bar for further studies, thus opening perspectives toward more reliable and accessible healthcare solutions worldwide.

## 6. Conclusion

The current study introduces NASNet-ViT, a hybrid deep learning model for reliably classifying lung disorders, such as lung cancer, COVID-19, TB, and normal pneumonia. The suggested model incorporates NASNet's convolutional feature extraction capabilities with the global attention mechanism capabilities of ViT to effectively handle the complexities in medical image analysis. Advanced pre-processing techniques, such as MixProcessing, enhanced the model's complex medical image processing ability. The proposed NASNet-ViT model produced remarkable metrics of 98.9% accuracy, sensitivity of 0.99, and specificity of 0.985, hence outperforming the current state-of-the-art architectures such as ResNet50, MobileNet, and MixNet-LD. It yields a highly efficient and scalable model with a size of 25.6 MB and computational time of 12.4 seconds, hence deployable in real-time in resource-constrained clinical environments. The study further points out that more validation on diverse multi-regional datasets must be done to have better generalizability. Future work could be directed toward integrating explainable AI techniques in order to build more trust among clinicians or exploring multimodal data to increase diagnosis capability. NASNet-ViT represents the state-of-the-art in lung disease diagnosis that is simultaneously robust, efficient, and scalable and closes the gap from advanced AI models to practical healthcare applications. It provides the starting point for further research into new ways of carrying out medical image analyses, benefiting both patient outcomes and clinical support worldwide.

**Data Availability Statement:** The data presented in this study are available in Kaggle at https://www.kaggle.com/datasets/omkarmanohardalvi/lungs-disease-dataset-4-types (accessed on 28 July 2023) and Chest CT-Scan Images Dataset. Available online: www.kaggle.com/datasets/mohamedhanyyy/chest-ctscan-images (accessed on 30 August 2023).

**Conflicts of Interest:** The authors of this work state that they have no conflicts of interest with relation to its publication.

**Ethics approval and consent to participate:** This study utilized publicly available datasets that are fully anonymized. Since no human subjects were directly involved, institutional review board (IRB) approval and informed consent were not required.

**Consent for publication:** Not applicable.

**Competing interests:** The authors declare no competing interests.